\documentclass[aps,prl,twocolumn,superscriptaddress,showpacs]{revtex4-1}

\usepackage{amsmath}
\usepackage{amsfonts}
\usepackage{amssymb}
\usepackage{graphicx}
\usepackage{bm}
\usepackage{color}

\begin{document}

\newcommand{\Fref}[1]{Figure~\ref{#1}~}
\newcommand{\fref}[1]{Fig.~\ref{#1}~}
\newcommand{\etal}{\emph{et al.}~}
\newcommand{\tc}{$T_c$~}
\newcommand{\spm}{$s_{\pm}$}~
\newcommand{\rr}{\textcolor{red}{\textbf{[ref?]}}}
\newcommand{\srp}{SrFe$_2$(As$_{1-x}$P$_{x}$)$_2$~}

\title{Nodal superconductivity in isovalently substituted SrFe$_2$(As$_{1-x}$P$_{x}$)$_2$  pnictide superconductor at the optimal doping, $x=$0.35}

\author{J.~Murphy}
\affiliation{The Ames Laboratory and Department of Physics and Astronomy, Iowa State University, Ames, IA 50011, USA}

\author{C.~P.~Strehlow}
\ 
\affiliation{The Ames Laboratory and Department of Physics and Astronomy, Iowa State University, Ames, IA 50011, USA}

\author{K.~ Cho}
\affiliation{The Ames Laboratory and Department of Physics and Astronomy, Iowa State University, Ames, IA 50011, USA}

\author{M.~A.~Tanatar}
\affiliation{The Ames Laboratory and Department of Physics and Astronomy, Iowa State University, Ames, IA 50011, USA}

\author{N.~Salovich}
\affiliation{Loomis Laboratory of Physics, University of Illinois at Urbana-Champaign, 1110 West Green St., Urbana, IL 61801}

\author{R.~W.~Giannetta}
\affiliation{Loomis Laboratory of Physics, University of Illinois at Urbana-Champaign, 1110 West Green St., Urbana, IL 61801}

\author{T. Kobayashi}
\affiliation{Department of Physics, Graduate School of Science, Osaka University, Osaka 560-0043, Japan}

\author{S.~Miyasaka}
\affiliation{Department of Physics, Graduate School of Science, Osaka University, Osaka 560-0043, Japan}
\affiliation{JST, Transformative Research-Project on Iron Pnictides (TRIP), Tokyo 102-0075, Japan}

\author{S.~Tajima}
\affiliation{Department of Physics, Graduate School of Science, Osaka University, Osaka 560-0043, Japan}
\affiliation{JST, Transformative Research-Project on Iron Pnictides (TRIP), Tokyo 102-0075, Japan}

\author{R. Prozorov}
\email[Corresponding author: ]{prozorov@ameslab.gov}
\affiliation{The Ames Laboratory and Department of Physics and Astronomy, Iowa State University, Ames, IA 50011, USA}

\date{\today}

\begin{abstract}
Temperature-dependent London penetration depth, $\lambda(T)$, was measured in optimally - doped, $x=$0.35, as-grown ($T_c \approx$25~K, RRR=$\rho(300K)/\rho(T_c)$=4.5) and annealed ($T_c \approx$35~K, RRR=6.4) single crystals of \srp iron - based superconductor. Annealing increases the RRR and decreases the absolute value of the London penetration depth from $\lambda(0) = 300 \pm 10$~nm in as-grown sample to $\lambda(0) = 275 \pm 10$~nm. At low temperatures, $\lambda(T) \sim T$ indicating superconducting gap with line nodes. Analysis of the full - temperature range superfluid density is consistent with the line nodes, but differs from the simple single - gap $d-$wave. The observed behavior is very similar to that of BaFe$_2$(As$_{1-x}$P$_{x}$)$_2$, showing that isovalently substituted pnictides are inherently different from the charge - doped materials.

\end{abstract}


\maketitle

Superconductivity in the $AE$Fe$_2$As$_2$ ($AE$ = Ba, Sr or Ca) - based compounds can be induced by substitution of each constituent element in the formula \cite{Johnston2010,Canfield2010review122,Paglione2010review,Stewart2011,HHWenAnnRev2011}. In BaFe$_2$As$_2$ (Ba122), hole- (K substitution of Ba, BaK122) and electron- (e.g., Co substitution of Fe, BaCo122) doping produces superconductors with the maximum superconducting transition temperature, \tc of 38 K and 24 K, respectively. The isovalent substitution of As with P (BaP122) induces superconductivity with maximum \tc of 31~K \cite{Kasahara2010a}. 

Despite proximity to an identical ground state of the parent compound and similar values of \tc, superconductors produced by isovalent doping have a gap structure entirely different from that produced by hole or electron doping. The superconducting gap in charge - doped BaK122 and BaCo122 is full and isotropic at the optimal doping \cite{ProzorovKoganROPP2011,ReidSUST2012}. It becomes anisotropic upon departure from the optimal doping toward either end of the ``superconducting dome'' and even develops nodes at the extreme doping levels \cite{Cho2012,Cho2012a,Martin2010,Reid2010,Tanatar2010c,Maiti2011a,Kim2012}. In sharp contrast, the superconducting gap of isovalently substituted BaP122 reveals line nodes irrespective of the doping level \cite{Hashimoto2010,Suzuki2011,Hashimoto2012}.

To-date, the dominant theory of iron - based superconductors is based on the extended \spm model \cite{Mazin2008}, which originally predicted equal full isotropic gaps. There were many refinements of this idea to describe the numerous experiments that showed anisotropic and even nodal gaps \cite{Hirschfeld2011,Maiti2011,Maiti2011a,Chubukov2012,Khodas2012}. It was found that nodal and nodeless ground states are energetically very close to each other and the fine tuning of the inter- and intra- band interactions and band-structure parameters may shift the balance towards one of them. In addition, there are other important ingredients, such as the interplay of magnetism and superconductivity \cite{Parker2009,Fernandes2010,Maiti2012} and the influence of disorder \cite{Kemper2009,Efremov2011,Vavilov2011,Bang2012}.

Still, it seems that BaP122 is unique. Why does the superconducting gap of this material remain nodal irrespective of doping level? Is its uniqueness linked to the high purity as suggested by the observation of quantum oscillations \cite{Carrington2011,Hashimoto2012}? This raises the question of whether other isovalently substituted 122 materials also have a nodal gap at optimal doping and if so, how do their properties compare to BaP122?

Superconductivity in SrFe$_2$(As$_{1-x}$P$_x$)$_2$ with $T_c = 27$~K at the optimal doping of $x=0.35$ was discovered in 2010 \cite{Shi2010}. Soon thereafter, single crystals with $T_c = 30$~K were obtained \cite{Kobayashi2011}. Currently, optimized annealing protocol results in high quality crystals with $T_c = 35$~K \cite{Nakajima2012}. The first studies of SrP122 have already shown it to be an unconventional superconductor. Specific heat and NMR studies are consistent with the nodal small gap and nodeless larger gaps \cite{Dulguun2012}. Microwave measurements find fractional power law behavior of the London penetration depth. These studies, together with an analysis of flux flow resistivity, suggested a nodal gap in one band and strong anisotropy in the other \cite{Takahashi2012} and this behavior was contrasted to that of BaP122.

In this Letter we present high - resolution measurements of London penetration depth in as-grown and annealed single crystals of \srp superconductor. 
The low temperature behavior suggests a superconducting gap with vertical line nodes, similar to the cuprates.  However, the full temperature dependence of the superfluid density shows that the gap structure is more complicated than a simple single band $d-$wave and may be more consistent with a gap with accidental nodes (see, e. g., \cite{Hirschfeld2011}).
The absolute value of the London penetration depth at the optimal doping decreases upon annealing from $\lambda(0) = 300\pm 10$ nm in as-grown samples to $\lambda(0) = 275\pm 10$ nm. Importantly, - the overall nodal behavior of \srp was found to follow closely that of BaP122.

\begin{figure}[tb]
\includegraphics[width=1.0\linewidth]{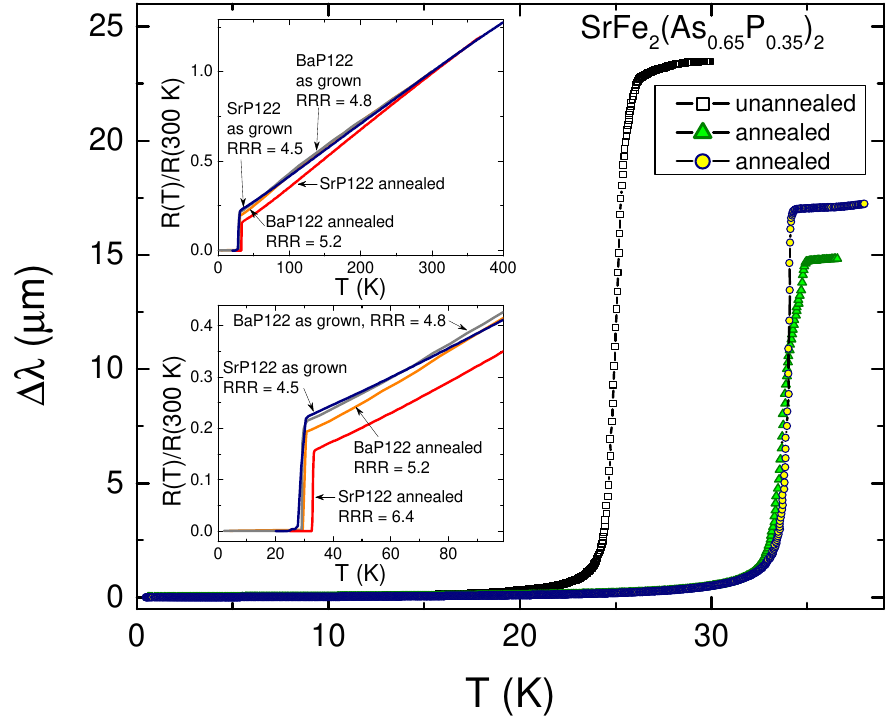}%
\caption{
Main pane: In-plane London penetration depth in single crystals of \srp, $x=$0.35, in the full temperature range showing one as-grown and two annealed samples. Top inset shows normalized resistivity, R($T$)/R(300~K) for as-grown and annealed SrP122 (this work) and BaP122 (Ref.~\onlinecite{Nakajima2012}). Lower inset shows the same data zoomed on in the vicinity of $T_c$.
}
\label{fig:lambda}
\end{figure}

Single crystals of SrFe$_{2}$(As$_{1-x}$P$_{x}$)$_{2}$ were grown using the self-flux method \cite{Kobayashi2011}. Samples were characterized by x-ray, magnetization and transport measurements and the composition was determined using EDX analysis, which yielded $x=$0.35. For London penetration depth measurements samples were selected from different batches by measuring the transition curves and finding the sharpest transition. The best samples were cut to a typical sample size of 0.5 $\times$ 0.5 $\times$ (0.02-0.1) mm$^3$. Annealing was shown to improve \tc from 31 K to 34.8 K and to increase the residual resistivity ratio, RRR=R(300~K)/R($T_c$) from 4.5 to 6.4. A lesser increase of \tc from 30 K to 31 K and of RRR from 4.8 to 5.2 after the annealing was reported for close to the optimal doping ($x=$0.32) BaP122 samples \cite{Nakajima2012}, see insets in Fig.~\ref{fig:lambda}. Furthermore, if we extrapolate linearly the resistivity curves to $T=0$, we obtain RRR(0) = 10.2 and 15.1 for as-grown and annealed SrP122, and RRR(0) = 7.1 and 8.1 for as-grown and annealed BaP122, respectively. By these measures, SrP122 appears to be cleaner than BaP122.

London penetration depth was measured by using a tunnel diode resonator (TDR) technique (see Ref.~\cite{Prozorov2006} and references therein). A tunnel diode supports self - oscillations in a high stability $LC$ tank circuit at the resonant frequency $f_0=1/2\pi\sqrt{LC}\approx$ 14 MHz in our setup. The sample is mounted on a movable sapphire rod whose temperature can be controlled independently of the coil and oscillator components. The excitation magnetic field of the coil is $\approx$ 20 mOe, well below first critical field. Insertion of the sample causes a frequency shift by an amount $\Delta f(T)=-G4\pi\chi(T)=G[1-(\lambda(T)/R)\tanh(R/\lambda(T))]$, which allows us to measure the change in London penetration depth, $\Delta\lambda(T)=\lambda(T)- \lambda(0)$ with a resolution of nearly 1 \AA. The geometric constant $G$ depends on the coil and sample volumes, demagnetization and empty coil resonance frequency and is measured directly by extracting the sample from the inductor coil while at base temperature \cite{Prozorov2000}. 

Figure~\ref{fig:lambda} shows the full temperature range variation of the in-plane London penetration depth, $\Delta\lambda(T)$, measured in an as - grown ($T_c = 27$~K) and two annealed ($T_c = 34.8$~K) single crystals of \srp, $x=$0.35. The insets show normalized resistivity,R($T$)/R(300~K) for as-grown and annealed SrP122 (this work) and BaP122 (Ref.~\onlinecite{Nakajima2012}) samples. Lower inset shows the data zoomed in the vicinity of $T_c$. Overall, the resistivity curves for SrP122 and BaP122 are virtually the same showing clear deviation from the Fermi liquid $T^2$ dependence at all temperatures, indicating proximity to the quantum critical point at the optimal doping \cite{Nakai2010,Carrington2011,Hashimoto2012,Iye2012,Kobayashi2011}.

\begin{figure}[tb]
\includegraphics[width=1.0\linewidth]{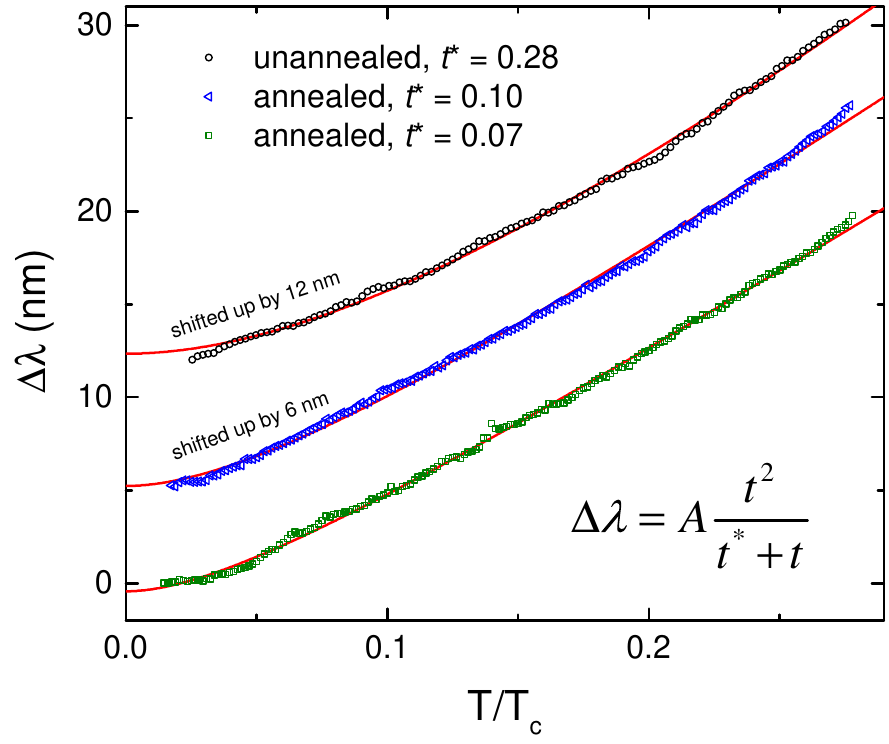}%
\caption{
Low - temperature part of $\Delta \lambda(T)$ for the three samples of \srp, $x=$0.35. Solid lines are the best fits to the Hirschfeld - Goldenfeld model \cite{Hirschfeld1993}, see text for discussion.
}
\label{fig:lambdaLT}
\end{figure}

Figure~\ref{fig:lambdaLT} shows the low temperature behavior of the penetration depth for three samples of \srp. Two of the curves are shifted vertically by 0.06 and 0.12 nm to avoid overlap. The linear temperature dependence is evident. Some rounding off at the low temperatures is due to impurity scattering as was shown for nodal cuprate superconductors by Hirschfeld and Goldenfeld  \cite{Hirschfeld1993}. Within their model the behavior at low temperatures can be approximated by $\Delta \lambda(T) = At^2/(t^*+t)$ where $t^*$ is a crossover temperature scale determined by unitary limit impurity scattering. Solid red curves in Fig.~\ref{fig:lambdaLT} show best fits to the data resulting in the crossover temperatures $t^* = $0.068, 0.101 and 0.285 for the three curves from bottom up. The amplitude $A$ also increases from the bottom to the top curve, $A=$88, 97 and 130 nm, respectively. A straightforward interpretation is that we are dealing with samples with different degrees of scattering from the cleanest (lowest curve) to the dirtiest (top curve) and such assignment is in line with the effect of annealing on resistivity and \tc. Good quality fits to the Hirschfeld - Goldenfeld formula, as shown in Fig.~\ref{fig:lambdaLT}, would appear to indicate the presence of line nodes. However, this is not sufficient for the determination of the topology of the nodal lines on the multi-band warped Fermi surface. 

For a full analysis we must determine the superfluid density over the entire temperature range. Knowing the variation of $\Delta \lambda (T)$, the superfluid density is given by $\rho_s(T)=\lambda^2(0)/\lambda^2(T)=(1+\Delta \lambda(T)/\lambda(0))^{-2}$, so we need to know the absolute value of zero - temperature penetration depth, $\lambda(0)$. To obtain this value we used TDR measurements of Al coated samples \cite{Prozorov2000a}. After initial measurement of $\Delta \lambda(T)$ each sample is uniformly coated with Al using magnetron sputtering and then remeasured \cite{Prozorov2000a,Gordon2010}. To ensure a uniform Al film thickness the sample is suspended by a fine wire from a rotating stage inside the sputter deposition chamber. The thickness of the Al layer, $d$, was measured using focused - ion beam cross-sectioning and imaging in SEM \cite{Gordon2010}. In our case $d=$73~nm is greater than the Al London penetration depth, $\lambda^{Al}(0) = 52$~nm. At $T<T^{Al}_c$, the effective penetration depth is given by:

\begin{equation}
\lambda_{eff}(T)=\lambda^{Al}(T)\frac{\lambda(T)+\lambda^{Al}(T)\tanh{\frac{d}{\lambda^{Al}(T)}}}
{\lambda^{Al}(T)+\lambda(T)\tanh{\frac{d}{\lambda^{Al}(T)}}}
\label{eq:lambda0}
\end{equation}

\noindent where $\lambda(T)$ is the London penetration depth of the material of interest. When Al becomes normal at $T^{Al}_c \approx 1.28$~K, $\lambda_{eff}(T)=d-\lambda(T^{Al}_c)$. Extrapolation of $\Delta \lambda(T)$ to $T=0$ shows that $\lambda(T^{Al}_c) \approx \lambda(0) +0.7$~nm and by using the BCS s-wave form of $\lambda(T)$ for Al, we can estimate the difference,$L = \lambda(0) - \lambda_{eff}(0)$. Solving numerically Eq.~\ref{eq:lambda0} we obtain $\lambda(T)$. Considering all the uncertainties, we estimate the accuracy as $\pm 10$~nm.

\begin{figure}[tb]
\includegraphics[width=1.0\linewidth]{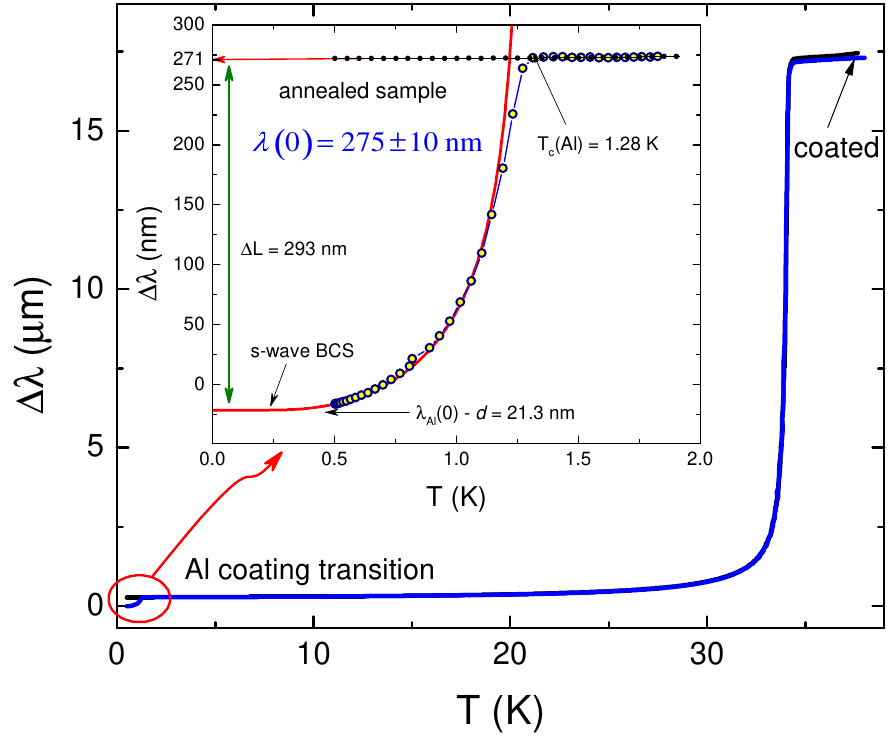}%
\caption{ Main panel: full - temperature $\Delta \lambda(T)$ of the same sample before and after aluminum coating showing that the curves are indistinguishable for $T>T^{Al}_c$. Inset shows the region of the Al transition. The curves are offset vertically by $\lambda_{eff}(T)=d-\lambda(T^{Al}_c)$ providing a rough visual estimate of $\lambda(0) \approx 271$~nm upon extrapolation of the uncoated sample curve to $T=0$. Numerical solution of Eq.~\ref{eq:lambda0} gives $\lambda(0) \approx 275 \pm 10$~nm.}
\label{fig:L0}
\end{figure}

Figure~\ref{fig:L0} illustrates the procedure to estimate the absolute value of $\lambda(0)$. Main panel shows full temperature - range $\Delta \lambda(T)$ for the same annealed sample measured before and after aluminum coating. Evidently, the curves reproduce each other perfectly for $T>T^{Al}_c$ indicating a good repeatability and stability of our measurements. The low - temperature part in the vicinity of the superconducting transition of the aluminum layer is shown in the inset in Fig.~\ref{fig:L0}. The curves are offset vertically, so that BCS extrapolation (shown my the solid line) to $T=0$ gives effective penetration depth of $\lambda_{eff}(T)=d-\lambda(T^{Al}_c) = 21.3$ nm. The difference between the uncoated sample and the coated sample at $T=0$ gives a rough visual estimate of $\lambda(0) = 271$~nm and the numerical solution of Eq.~\ref{eq:lambda0} (with the discussed above uncertainty of 10 nm) finally gives  $\lambda(0) \approx 275 \pm 10$~nm. 
Applying the same procedure, we obtained $\lambda(0)=300 \pm 10$~nm for the as-grown sample, consistent with the assumption of an enhanced pair - breaking compared to the annealed samples. In BaP122 at the optimal doping, $x=$0.30, we obtained a comparable magnitude of  $\lambda(0) \approx 330$~nm, but the situation is complicated by the strong doping dependence of $\lambda(0)$ due to the quantum critical point hidden beneath the dome \cite{Hashimoto2012}. Whether the same features exist in SrP122 requires a systematic doping study.

Combining the results presented in Fig.~\ref{fig:lambdaLT} and Fig.~\ref{fig:L0} we can compare the rate of change of the penetration depth with temperature observed in other clean nodal superconductors with the current work. In a $d-$wave superconductor with vertical line nodes, the amplitude of the (linear) low - temperature variation of the penetration depth is given by \cite{Xu1995}: 

\begin{equation}
\frac{d\left( \lambda /\lambda (0)\right) }{d\left( T/T_{c}\right) }\equiv \frac{d\lambdabar }{dt}=\frac{2\ln 2}{\left( d\Delta /d\varphi \right) _{\varphi \rightarrow node}}
\label{eq:dLdT}
\end{equation}

\noindent where $\left( d\Delta /d\varphi \right) _{\varphi \rightarrow node}$ is the slope of the angle - dependent superconducting gap approaching the node position on the Fermi surface. In the case of $d-$wave pairing, $\Delta (\varphi) =\Delta (0)\cos \left( 2\varphi \right)$ and $d\lambdabar /dt=T_{c}\ln 2/\Delta (0)=\ln 2/2.14=0.32$.
For YBCO, the measured $d\lambdabar /dt=0.33$ \cite{Zhang1994,Prozorov2000a} and for BSCCO2212 the observed value is $d\lambdabar /dt=0.39$ \cite{Jacobs1995,Prozorov2000a}, - both are quite close to the theoretical prediction. In the present case of \srp, we obtained $d\lambdabar /dt=0.28$. For comparison, in BaP122, $d\lambdabar /dt=$0.42 and 0.38 for $x=0.30$ ($\lambda(0) = $330~nm) and $x=0.33$ ($\lambda(0) = $215~nm), respectively \cite{Hashimoto2012}. These values are in a reasonable agreement with the theoretical value of 0.32 showing that the node topology is not much different from a standard $d-$wave symmetry.


\begin{figure}[tb]
\includegraphics[width=1.0\linewidth]{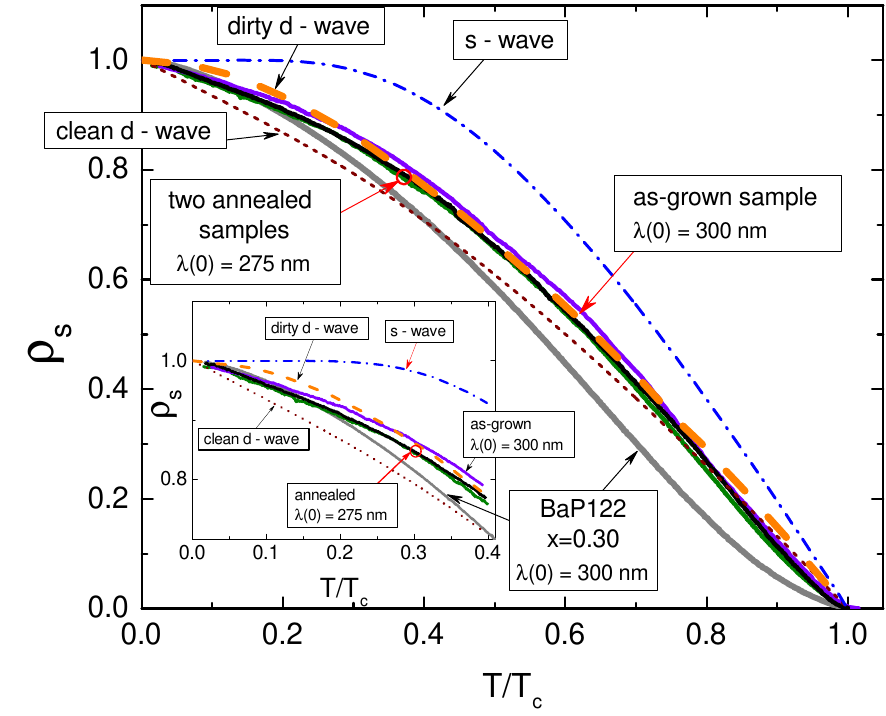}%
\caption{Comparison of the superfluid density, $\rho(T)$, for three samples of \srp with the prediction of a two - dimensional $d-$wave pairing (short-dashed line - clean and dashed line - dirty limits) and isotropic $s-$wave (dot-dashed line). We also show $\rho(T)$ for BaP122 (gray line, $x=$0.30, $\lambda(0)=$330~nm). (Inset) Expanded view of low temperature region.}
\label{fig:SFD}
\end{figure}

Figure~\ref{fig:SFD} shows experimental superfluid densities constructed with the estimated values of $\lambda(0)$. The data are compared with the expectations for $d-$wave pairing (short-dashed line - clean and dashed line - dirty limits) and isotropic $s-$wave (dot-dashed line). The data are in a complete disagreement with the exponentially saturating $s-$wave curve. Instead, the data show a clear $T-$linear variation at low temperatures. For comparison, the data for BaP122 are also shown by the gray line. The curves for BaP122 and SrP122 overlap at the low temperatures (below 0.2 \tc, see inset), but deviate at higher temperatures. This difference must be due to the difference in the gap magnitudes and anisotropies in these multi-gap systems, but the low - temperature behavior is determined by the nodal quasiparticles and the similarity of the data implies that the nodal structure of SrP122 and BaP122 is similar. The deviation from the 2D $d-$wave could be due to geometry of the nodal lines, - perhaps forming the loops in the electron bands \cite{Carrington2011,Graser2010,Maiti2010,Shimojima2012,Suzuki2011,Hirschfeld2011}.

In conclusion, measurements of the London penetration depth, $\lambda(T)$, in optimally - doped as - grown and annealed single crystals of \srp iron - based superconductor provide clear evidence for line nodes. The absolute value of London penetration depth decreases with annealing from $\lambda(0) = 300 \pm 10$~nm to $\lambda(0) = 275 \pm 10$~nm. The slope $d\lambdabar /dt=0.28$ is consistent with the expectations for the superconducting gap with line nodes,  $d\lambdabar /dt=\ln 2/2.14=0.32$ and comparable to the measured values in YBCO and BSCCO2212. The superfluid density $\rho(T)$ differs from the prediction for the vertical infinite line nodes (as in a simple single - band $d-$wave) and requires an analysis withing a full three - dimensional band-structure. Overall, our results indicate that \srp behaves very similarly to BaFe$_2$(As$_{1-x}$P$_{x}$)$_2$ both from transport and superfluid response points of view and it seems that isovalently substituted pnictides are inherently different from the charge - doped materials.

The work at Ames was supported by the U.S. Department of Energy, Office of Basic Energy Sciences, Division of Materials Sciences and Engineering under contract No. DE-AC02-07CH11358. Work at UIUC was supported by the Center for Emergent Superconductivity, an Energy Frontier Research Center funded by the US Department of Energy, Office of Science, Office of Basic Energy Sciences under Award No. DE-AC0298CH1088. Work at Osaka was partly supported by a Grant-in-Aid IRON-SEA from the Japan Science and Technology Agency (JST).

%

\end{document}